\documentstyle[prb,aps,multicol,epsfig]{revtex}
\def \be{\begin{equation}}
\def \ee{\end{equation}}

\def \D{{\cal D}}
\def \C{{\cal C}}
\def \Nch{{N_{\rm ch}}}

\def \mls{\delta_1}
\renewcommand{\narrowtext}{\begin{multicols}{2} \global\columnwidth20.5pc}
\renewcommand{\widetext}{\end{multicols} \global\columnwidth42.5pc}
\newcommand{\Lrule}{\vspace*{-0.2in}\noindent\vrule width3.5in height.2pt
  depth.2pt \vrule depth0em height1em}
\newcommand{\Rrule}{\vspace{-0.1in}\hfill\vrule depth1em height0pt \vrule
  width3.5in height.2pt depth.2pt\vspace*{-0.1in}}

\begin{document}

\bibliographystyle{simpl1}

\title{Conductance Fluctuations of Open Quantum Dots under Microwave
Radiation.}  
\date{\today} 

\author{Maxim~G.~Vavilov$^a$ and Igor
L.~Aleiner$^b$} 

\address{ $^a$ Laboratory of Atomic and Solid State
Physics, Cornell University, Ithaca, NY 14853\\ $^b$ Department of
Physics and Astronomy, SUNY at Stony Brook, Stony Brook, NY 11794 }
\maketitle
\begin{abstract}
We develop a time dependent random matrix theory describing the
influence of a time-dependent perturbation on mesoscopic conductance
fluctuations in open quantum dots. The effect of external field is
taken into account to all orders of perturbation theory, and our
results are applicable to both weak and strong fields. We obtain
temperature and magnetic field dependences of conductance
fluctuations. The amplitude of 
conductance fluctuations is determined by electron temperature in
the leads rather than by the width of electron distribution function
in the dot. The asymmetry of conductance with respect to inversion of
applied magnetic field is the main feature allowing to distinguish
the effect of direct suppression of quantum interference from the simple 
heating if the frequency of external radiation is larger than 
the temperature of the leads $\hbar\omega \gg T$.
\end{abstract}
\draft

\narrowtext

{\em Introduction} --
Transport coefficients of disordered and chaotic 
electron systems fluctuate from sample to
sample \cite{A1,LS,A,B,AK,H1,Huibers}. 
These fluctuations are commonly called mesoscopic
fluctuations. Mesoscopic fluctuations of conductance of
non-interacting systems are universal. The universality means
that the variance of the conductance $\langle \delta g^2\rangle$ is of
the order $G_0^2$,  where $G_0=e^2/\pi\hbar$ is the quantum of
conductance and is weakly dependent of the sample geometry, see \cite{B,AK}.

The fluctuations of transport properties of electron systems is a
quantum mechanical phenomenon based on the interference of quantum
states. As any other interference phenomena, conductance fluctuations
are very sensitive to inelastic processes, commonly referred to as
dephasing \cite{AAKL}. The dephasing processes in open quantum dots
were considered on the purely phenomenological basis \cite{BM1}.
First microscopic consideration of the microwave radiation on the weak
localization in quantum dots was performed in Ref.~\cite{VA}.
In this reference, the concept of the time dependent random matrix
theory (TRMT) was used.

The purpose of the present paper is to extend the results of
Ref.~\cite{VA} to describe the effect of the external microwave
radiation on the mesoscopic fluctuations of the conductance. The
ultimate goal is to identify observable features which allow one to
distinguish the effect of the external field to the dot itself from a
simple heating  \cite{Huibers}.

However, there is a significant difference in the calculation of the
mesoscopic conductance fluctuations and the averaged conductance
\cite{VA}. Because, the dot is subjected to the external classical 
radiation which produces non-equilibrium in the dot, the 
d.c.-current $I_0$ through the dot is finite (though randomly changing
from one configuration to another) even if the {\em d.c.}-voltage 
$V=V_{\rm l}-V_{\rm r}$ across the
dot is zero, see Fig.~1.
 This current $I_0$ is due either to the photovoltaic effect (see
\cite{VAA} and references therein) or to rectification of a.c.-bias across the
dot, see \cite{Brouwer-ac}. We are interested in the linear response to
the applied {\em d.c.}-voltage across the dot:
\begin{equation}
\label{16}
I_{\rm dc} = I_{0}+gV+{\cal O}(V^2).
\end{equation}
In principal, the linear in $V$ contribution to the current comes 
comes from two sources:
({\it i}) the non-equilibrium of the distribution functions in the 
leads, ({\it ii}) change in the photovoltaic current,
correspondent to a different realization of the dot due to the finite
bias.
Nevertheless, we will show that due to the electro-neutrality condition
the non-equilibrium current prevails \cite{Y}.

Closing the introductory part, we note that the
present paper has a certain overlap with the recent preprint by Wang
and Kravtsov \cite{WK}, where the conductance fluctuations were
calculated for open quantum dots subjected to a periodic $ac$
pumping.  Our treatment is different in several aspects. Firstly, our
results are applicable for the frequencies of the external radiation
$\omega$ smaller than the Thouless energy of the dot, $E_T$, whereas
treatment of Ref.~\cite{WK} is valid in the opposite regime.
Secondly, unlike Ref.~\cite{WK}, we will restrict ourselves to the
case of the monochromatic radiation acting on the dot.  Finally, we
will highlight the role of the electro-neutrality requirement in a
separability of the photovoltaic effect and the mesoscopic conductance
fluctuations, which was not done in Ref.~\cite{WK}.

{\em Model}---
We apply the random matrix theory (RMT) 
to study the conductance of open quantum dots, see Ref.~\cite{B}. All 
corrections to the RMT are governed by a small parameter
$N_{\rm ch}/g_{\rm dot}$, where 
$g_{\rm dot}=E_{\rm T}/\delta_1$ and $\delta_1$ is
the mean level spacing, see Ref.~\cite{ABG} for the detailed discussion. 
We consider the conductance fluctuations of quantum 
dots with a large number of open 
channels $N_{\rm ch}\gg 1$. In this approximation, we neglect the effects 
of the electron--electron interaction on the conductance which are as
small 
as $1/N_{\rm ch}^2$\cite{ABG,Brouwer}, while the 
conductance fluctuations are proportional
to $1/N_{\rm ch}$. The same condition also  allows us to use a
conventional diagrammatic technique\cite{AGD} to 
take the ensemble average. External microwave radiation is
modeled as time dependent random part of the Hamiltonian of the dot.

\begin{figure}
\centerline{\psfig{figure=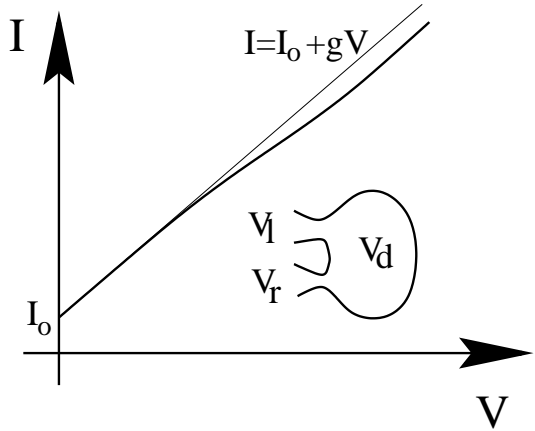,width=7cm}}
{
\caption{
An open quantum dot is connected to two leads with applied voltages
$V_{\rm l,r}$. The measured current through the
dot has an offset $I_0$ at zero bias  and the linear
response to small applied voltage $V=V_{\rm l}-V_{\rm r}$. }
}
\end{figure}

The Hamiltonian of the system is
\cite{ABG}:
\be
\label{1}
{\cal H}(t)={\cal H}_{\rm D}(t)+{\cal H}_{\rm L}+{\cal H}_{\rm LD},
\ee
where $\hat H_{\rm D}$ is the Hamiltonian of the electrons in the dot,
which is determined by the $M\times M$ matrix $H_{nm}$:
\be
\label{2}
{\cal H}_{\rm D}(t)=\sum_{n,m=1}^M \psi^\dag_n H_{{\rm D}nm}(t) \psi_m
+E_c{n}^2,
\ee
$\psi_n$ corresponds to the states of the dot and
the thermodynamic limit $M \to\infty$ is assumed,  $E_c$ is the
charging energy, and ${n}=\sum_{m=1}^M  \psi^\dag_m \psi_m$,
the last term in Eq.~(\ref{2}) is the largest contribution to
the interaction effects in quantum dot, see Ref.~\cite{ABG}
for the discussion of the status of this approximation. 
Matrix $\hat H_{\rm D}(t)$ is given by
\be
\label{03}
\hat H_{\rm D}(t)=\hat H+\hat V \varphi(t).
\ee
Here the time independent part of the Hamiltonian $\hat H$ is a
random realization of a $M\times M$ matrix, which obeys the correlation
function 
\begin{eqnarray}
\label{04}
&&\langle H_{nm}(\Phi_1) H^*_{n'm'}(\Phi_2)\rangle=
M\left(\frac{\mls}{\pi}\right)^2 \\
&&\quad
\times
\left[\left(1-\frac{N_{\rm d}}{4M}\right)
\delta_{nn'}\delta_{mm'}
+\left(1-\frac{N_{\rm c}}{4M}\right)\delta_{mn'}\delta_{nm'}\right], 
\nonumber
\end{eqnarray}
where $\mls$ is the mean level spacing of the dot and
parameters $N_{d,c}$ describe the effect of the magnetic field on the
dot\cite{ABG}.
These parameters can be estimated as $N_{\rm d,c} \simeq g_{\rm dot} 
\left(\Phi_1\mp\Phi_2\right)^2/\Phi_0^2$
where $\Phi_{1,2}$ is the magnetic flux through the dot and $\Phi_0=hc/e$ is
the flux quantum. The time dependent perturbation is described by
symmetric $M\times M$ matrices $V_{nm}$ and function $\varphi(t)$
of time $t$.
We assume that the perturbation is harmonic with single frequency
$\omega$, $\varphi (t)=\cos \omega t$, even though most of the
consideration [up to Eq.~(\ref{24})] is valid for an arbitrary function.
The effect of the perturbation on   
the system is totally determined by two parameters, \cite{fn1}
\be
Z= \frac{1}{M}{\rm Tr} \hat{V}, \quad
C_0 = \frac{\pi}{M^2\mls}{\rm Tr} \hat{V}^2. 
\label{05}
\ee 
Parameter $Z$ has a meaning of the average velocity of the energy
levels of the dot under the external perturbation and can be omitted
from our consideration due to screening [see below]. The
parameter $C_0$ characterizes its typical deviation, \cite{SA}. 
Since all the physical responses of the system are characterized by
the same parameters, the value of $C_0$ can be eliminated by an independent
measurement.

The  electron spectrum in the leads near Fermi
surface can be linearized:
\be
\label{9}
{\cal H}_L=\hbar v_{\rm F}\sum_{\alpha, k} k \psi^\dag_\alpha(k)\psi_\alpha(k),
\ee
where  $\psi_{\alpha}(k)$ denotes different electron states in the
leads, $k$ labels the continuum of momentum states in each 
channel $\alpha$, $\hbar v_{\rm F}=1/2\pi \nu$ is the
Fermi velocity and $\nu$ is the density of states per channel at the
Fermi surface. We put $\hbar =1$ in all intermediate formulas below. 

The coupling between the dot and the leads is
\be
\label{7}
{\cal H}_{\rm LD}=\sum_{\alpha, n, k}\left( W_{n \alpha}\psi^\dag_\alpha
(k)
\psi_n+
{\rm H.c.}\right).
\ee
For the reflectionless point contacts, 
the coupling constants, $W_{n \alpha}$,
in Eq.(\ref{7}) are given by \cite{B,ABG}:
\be
\label{8}
W_{n \alpha}=\cases{\displaystyle
\sqrt{\frac{M\mls}{\pi^2\nu}}, & if $n=\alpha\leq N_{\rm ch}$,\cr
0,& otherwise.
}
\ee

For open dots with a large number of open channels $N_{\rm ch}\gg 1$ 
the interaction term can be
treated within mean field approximation, so that the Hamiltonian
(\ref{2}) takes the form
\begin{mathletters}
\begin{eqnarray}
{\cal H}_{\rm D}^{\rm mf}(t)& = & \sum_{n,m=1}^M \psi^\dag_n 
\left[H_{{\rm D}nm}(t)+ e V_d(t) \delta_{nm}\right]
\psi_m, 
\label{99} 
\\
e V_d & = & 2 E_c\langle {n} \rangle_{q},
\label{100}
\end{eqnarray}
\end{mathletters}
where $\langle {n} \rangle_{q}$ stands for the quantum mechanical
(but not ensemble) of the number of electrons in the dot.
Corrections to mean-field treatment (\ref{99}) were calculated in
Ref.~\cite{Brouwer} and shown to be small as $1/N_{\rm ch}^2$.

In the mean field approximation (\ref{99}), one can introduce one-particle
$S$ - matrix ${\cal S}_{\alpha\beta}(t_1,t_2)$ as
\begin{eqnarray}
\label{13}
\label{S}
{\cal S}_{\alpha\beta}(t,t')=
\delta_{\alpha\beta}\delta(t-t')-2\pi i\nu
W^\dag_{\alpha n} G^{({\rm R})}_{nm}(t,t') W_{m\beta},
\end{eqnarray}
and the Green functions $\hat G^{({\rm R,A})}(t,t')$ are the solutions of:
\begin{eqnarray}
\nonumber
& & 
\left(i\frac{\partial}{\partial t}  -  {\hat H}_{\rm D}(t)-eV_{\rm d}(t) 
\pm i\pi\nu \hat{W}\hat{W}^\dag \right)\hat{G}^{({\rm R,A})}(t,t')
\\
& & = \delta(t-t'),
\label{14}
\end{eqnarray}
where the matrices $\hat{H}_{\rm D}$ and $\hat{W}$ are
defined by Eqs.~(\ref{03}) and (\ref{8}).

The d.c.-current through the dot is given by, see \cite{VAA}:
\begin{eqnarray}
I_{\rm dc} & = & {e}\int_0^{T_{\rm p}}\!\!
\frac{dt}{T_{\rm p}} \int\! dt_1dt_2
{\rm Tr}\left\{ \hat f(t_1-t_2)
\right.
\nonumber
\\
& \times &
\left.
\left(
\hat{\cal S}^{\dagger}(t_2,t) \hat \Lambda
\hat{\cal S}(t,t_1) - \hat \Lambda \delta(t_2-t_1)
\right)
\right\}
\label{10}
\end{eqnarray}
where $T_p$ is the period of the external perturbation,
 $\hat{f}(t)$ is related to the 
Fourier transform of the electron distribution
function in the $\alpha$th channel as
\be
\label{11}
f_{\alpha\beta}(t)=\delta_{\alpha\beta}
\frac{i Te^{-ieV_\alpha t} }{\sinh\pi T t}
\ee
and
\begin{equation}
\label{12}
\Lambda_{\alpha\beta}=\delta_{\alpha\beta}\cases{\displaystyle 
\frac{N_{\rm r}}{N_{\rm ch}}, &
for $1\leq \alpha\leq N_{\rm l}$; \cr
\displaystyle -\frac{N_{\rm l}}{N_{\rm ch}}, &
for $N_{\rm l}+1\leq \alpha\leq N_{\rm ch}$. \cr
}
\end{equation}
The spin degeneracy is taken into account in Eq.~(\ref{10}). We assume
that the degeneracy is not lifted by magnetic field.

To complete the theory one needs an equation for the averaged 
number of particles $\langle {n} \rangle_{q}$, 
see Eq.~(\ref{100}). It is found from the continuity
relation as
\begin{eqnarray}
\frac{d \langle {n}(t) \rangle_{q}}{dt} &=& 
- \int\! dt_1dt_2 {\rm Tr}\left\{ \hat f(t_1-t_2) \right.
\label{110}
\\ 
& \times & 
\left.
\left(
\hat{\cal S}^{\dagger}(t_2,t) \hat{\cal S}(t,t_1) - \delta(t_2-t_1)
\right)
\right\}
\nonumber
\end{eqnarray}

Equations (\ref{10}) --- (\ref{110}) are similar to those used in
Ref.~\cite{Buttikker} for studying the frequency dependence of the
conductance of mesoscopic system.

{\em Ensemble averaging} ---
Our goal now is to perform calculations of the conductance correlation
function 
\be
\label{R}
{\cal R}(\Phi_1, \Phi_2)=\langle g(\Phi_1) g(\Phi_2)\rangle - \langle
 g(\Phi_1) \rangle\langle g(\Phi_2)\rangle, 
\ee
 using the model outlined above.

\begin{figure}
\centerline{\psfig{figure=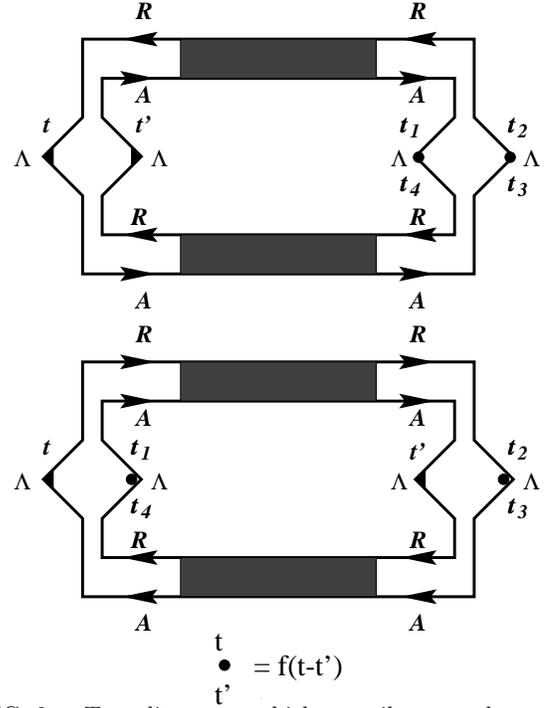,width=7cm}}
{
\caption{
Two diagrams, which contribute to the conductance correlation function
$R$. }}
\end{figure}

We use the leading approximation in small parameter $1/N_{\rm ch}$.
The fluctuations of the conductance are smaller than its average value
and we can use instead of sample specific Eq.~(\ref{110}) its ensemble
averaged counterpart:
\begin{eqnarray}
&&\frac{d \langle {n} (t) \rangle_{q}}{dt}
= -  \frac{\Nch\mls}{2\pi}\langle {n}(t) \rangle_{q}
\label{110a} \\
&& + \frac{eN_{\rm l}}{\pi}\left(V_{\rm l} - V_{\rm d} - 
Z\varphi(t)\right)
+ \frac{eN_{\rm r}}{\pi}\left(V_{\rm r} - V_{\rm d} - Z\varphi(t)\right)
.
\nonumber
\end{eqnarray}
Equation (\ref{110a}) is nothing but a discrete form of the
diffusion equation for the bulk system and the last two terms correspond to
the divergence of the drift current. Substituting Eq.~(\ref{100})
into Eq.~(\ref{110a}), solving the resulting differential equation,
we find
\begin{eqnarray}
e V_{\rm d}(t) + Z\varphi(t) &=&
\frac{4e E_{\rm c}}{4 E_{\rm c} + \mls}
\frac{N_{\rm l} V_{\rm l}+N_{\rm r} V_{\rm r}}
{\Nch} 
\label{200}\\
&+& Z 
\frac{\mls+ (2\pi/N_{\rm ch}) \partial_t}
{\mls+4 E_{\rm c} + (2\pi/N_{\rm ch})\partial_t }\varphi(t).
\nonumber
\end{eqnarray}
We notice from Eq.~(\ref{200}) that the characteristic energy
scale governing charge dynamics is $E_{\rm c}N_{\rm ch}/2\pi$. Usually,
this scale is of the order of the Thouless energy, $E_T$. Because
all the random matrix theory is capable to describe the energy scale
only smaller than $E_T$, we can consider only $\omega \ll
E_T \simeq E_{\rm c}N_{\rm ch}/2\pi$. Moreover, for the small quantum
dot $E_{\rm c} \gg \delta_1$, so that Eq.~(\ref{200}) gives
\be
V_{\rm d} = \frac{N_{\rm l} V_{\rm l}+N_{\rm r} V_{\rm r}}{\Nch}
\label{Vd}
\ee
and the time dependent perturbation (\ref{05}) 
can be considered as traceless,  $Z=0$.
This constant in time component of the bias of the dot
can be
removed from  Eq.~(\ref{14}) for the Green function 
by the following gauge transformation:
\begin{equation}
\label{gaugetransform}
\label{202}
\hat G(t,t')=\left. 
\hat G\right|_{V_{\rm d}=0}(t,t') e^{- i e V_{\rm d}(t-t')}.
\end{equation}

Substituting Eq.~(\ref{202}) into Eq.~(\ref{10}) and
expanding up to the first power in $V=V_{\rm l}-V_{\rm r}$,
we find 
\begin{eqnarray}
\label{17}
g  = \frac{\partial I_{\rm dc}}{\partial V}  = g_{\rm cl} & +&
\displaystyle
G_0
\int\limits_0^{T_{\rm p}}\!\!\! dt\!\! \int\limits_{-\infty}^{+\infty}
\!\!\!dt_1dt_2
F(T(t_1-t_2))
\\
\nonumber
&\times & {\rm Tr}
\Big{\{}
\hat {\cal S}(t,t_1)\hat \Lambda {\cal S}^\dagger(t_2,t)\hat \Lambda
\Big{\}}.
\end{eqnarray}
Here 
\[
g_{\rm cl}= G_0 \frac{N_{\rm l}N_{\rm r}}{N_{\rm ch}}
\] 
is the classical conductance of the dot, $G_0=e^2/\pi\hbar$ is the quantum
conductance and  $F(x)$ is the Fourier transform of the derivative of
electron distribution function:
\begin{equation}
\label{18}
F(x)=
\frac{\pi x}{\sinh\pi x}.
\end{equation}

The correlation function of the conductance 
(\ref{R}) is given by the diagrams
shown in Fig. 2 and can be found from the following analytical
expression 
\widetext
\Lrule
\begin{eqnarray}
\label{19}
R &  =&2 \frac{g_{cl}^2\mls^2}{4\pi^2}
\int\limits_0^{T_{\rm p}}\!\!
\frac{dtdt'}{T_{\rm p}^2}\int\limits_0^{\infty}\!\!
d\tau F^2(T\tau)
\int\limits_{\tau/2}^{\infty}\!\!d\theta 
\left[
\D\left(\frac{t+t'}{2}+\theta,\frac{t+t'}{2}-\frac{\tau}{2},t-t' \right) 
\D\left(\frac{t+t'}{2}+\theta,\frac{t+t'}{2}+\frac{\tau}{2},t-t'\right)
\right.
\nonumber
\\
& &+ \left. 
\C \left(
t-t'+\theta+\frac{\tau}{2},t-t'-\theta-\frac{\tau}{2},\frac{t+t'}{2}+
\frac{\tau}{4} 
\right)
\C
\left(
t'-t+\theta-\frac{\tau}{2},t'-t-\theta+\frac{\tau}{2},\frac{t+t'}{2}-
\frac{\tau}{4} 
\right)
\right].
\end{eqnarray}
\Rrule   
\narrowtext
The diffuson and the Cooperon are defined by the following equations:
\begin{mathletters}
\begin{eqnarray}
\label{20}
{\cal C}(\tau_1,\tau_2,t) & = & 
\Theta(\tau_1-\tau_2)
\exp\left(-\frac{1}{2}\int_{\tau_2}^{\tau_1}
K_{\rm c}(\tau,t)
d\tau \right), \\
\label{21}
{\cal D}(t_1,t_2,\tau) & = &\Theta(t_1-t_2) \exp\left(-\int_{t_2}^{t_1} 
K_{\rm d}(\tau,t) d t \right), 
\end{eqnarray}
\end{mathletters}
where 
\begin{mathletters}
\begin{eqnarray}
\label{22}
K_{\rm d,c} & = & \gamma_{\rm d,c} +
C_0\left(\varphi(t+\tau/2)-\varphi(t-\tau/2)\right)^2
\\
\displaystyle
\gamma_{\rm d, c} & = & \frac{\mls}{2\pi}
\left( N_{\rm ch}+N_{\rm d,c}\right),
\label{22.b}
\end{eqnarray}
\end{mathletters}
and parameters $N_{\rm d,c}$ describe the effect of the magnetic field,
see Eq.~(\ref{04}).

\begin{figure}
\centerline{\psfig{figure=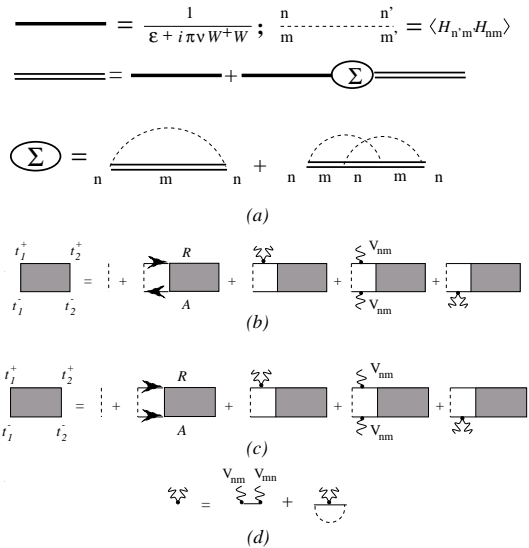,width=7cm }}
{
\caption{
The elements of the diagrammatic technique. }}
\end{figure}

Derivation of Eq.~(\ref{22.b}) deserves a little bit of additional
discussion.  One notices, that the diagrams of Fig.~3 does not contain
any piece corresponding to the classical distribution function in the
dot.  We can trace it into the expression for conductance (\ref{17})
which contains traceless vertices $\hat \Lambda$, which can not be
dressed by the dashed line. On the other hand, any vertex with
finite trace corresponds the modified distribution function of
electrons in the dot and represents the effect of heating ~\cite{VAA}.
Since the distribution function is not dressed in the expression for
conductance fluctuations, we conclude, that the effect of heating is
not relevant for the conductance fluctuations and the temperature
dependence of the conductance fluctuations is uniquely determined by
{\em the electron temperature in the leads\cite{Y}}.  That means that
contrary to the common believes, see e.g. Ref.~\cite{Huibers} the
amplitude of the mesoscopic fluctuations can {\em not} be used for the
study of the distribution function of electrons in the dot.  From the
theoretical side, it is important to emphasize, that the appearance of the
traceless vertices is determined solely by the electro-neutrality
condition (\ref{Vd}), any other choice of the dot bias would lead to
the change in the photovoltaic current ~\cite{VAA}.

{\em Limiting cases} --- 
Below we consider the limit of high ($\hbar\omega\gg C $) and low
($\hbar\omega\ll \gamma_{\rm d,c}$) frequencies. 
For the high frequencies, $\omega\gg C $, we
obtain 
\begin{equation}
\label{24}
R = \frac{\mls^2 g_{\rm cl}^2}{4\pi^2}\left(\frac{1}{\gamma_{\rm d}^2}
Q_{\rm d}
\left(
\frac{C_0}{\gamma_{\rm d}},
\frac{T}{\gamma_{\rm d}}
\right)
+
\frac{1}{\gamma_{\rm c}^2} Q_{\rm c}
\left(
\frac{C_0}{\gamma_{\rm c}},
\frac{T}{\gamma_{\rm c}},
\frac{\hbar\omega}{\gamma_{\rm c}}
\right)
\right),
\end{equation} 
where dimensionless $Q-$ functions are given by 
\begin{mathletters}
\begin{eqnarray}
\label{25}
Q_{\rm d}(x,y) & = &
\int\limits_0^\infty d\tau F^2(y\tau) \\
&\times &  
\int\limits_0^{2\pi} \frac{\exp\left(-\tau
\left(1+2x\sin^2\zeta/2\right)\right)}{1+2x\sin^2\zeta/2}\frac{d\zeta}{2\pi},
\nonumber
\\  
\label{26}
Q_{\rm c}(x,y,z) & = & \int\limits_0^\infty d\tau F^2(y\tau) \\
& \times & 
\int\limits_0^{2\pi}
\frac{\exp\left(-\tau \left(1+2x\sin^2\zeta/2\right)\right)}
{1+x(\sin^2\zeta/2+\sin^2(\zeta/2+z\tau/2 ))}\frac{d\zeta}{2\pi}.
\nonumber 
\end{eqnarray}  
\end{mathletters}

Let us now consider the dependence of the functions $Q_{\rm d,c}$ and 
$Q_{\rm c}(x,y,z)$ from Eq.~(\ref{26}) 
on temperature $y$. For the limit of high
temperature, $y \gg 1$, we obtain
\begin{equation}
\label{27}
Q_{\rm c}(x,y,z)\approx Q_{\rm d}(x,y)\approx 
\frac{\pi^2}{3 y}\frac{1}{\sqrt{1+2x}}. 
\end{equation}
The equality between functions $Q_{\rm c}$ and $Q_{\rm d}$ means 
the magnetic field symmetry of the conductance \cite{Onsager}. Indeed, using
Eqs.(\ref{22}), (\ref{24})
and (\ref{27}) we observe, that 
\be
R (\Phi_1,\Phi_2) =
R(\Phi_1,-\Phi_2).
\label{ons1}
\ee 
However, in low temperature limit $y \ll 1$, we obtain for $x\gg 1$
\begin{mathletters}
\begin{eqnarray}
\label{28}
Q_{\rm d}(x,0)=\frac{1+x}{(1+2x)^{3/2}} & \approx & \frac{1}{2\sqrt{2x}}.\\
\label{28a}
Q_{\rm c}(x,0,z)& \approx & \frac{1}{2x}.
\end{eqnarray}
\end{mathletters}

\begin{figure}
\centerline{\psfig{figure=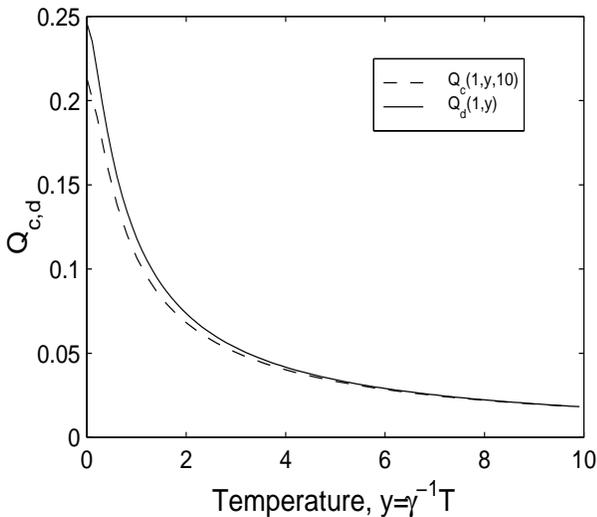,width=8cm,height=7cm}}
\caption{
Functions $Q_c(x,y,z)$ and $Q_d(x,y)$ computed for $x=1$ and
$z=10$. As temperature $y=T/\gamma$ increases, function $Q_{\rm c}(x,y,z)$
approaches frequency independent function $Q_{\rm d}(x,y)$. }
\end{figure}

Comparison of Eqs.(\ref{22}), (\ref{24}) and (\ref{28}) reveals
an important fingerprint of the dephasing by the external radiation
--- violation of the
Onsager relation
\be
\frac{R (\Phi_1,-\Phi_2)}{R(\Phi_1,\Phi_2)}
=
\sqrt{\frac{2\tilde \gamma}{C_0}}
,
\label{ons2}
\ee
where $\tilde\gamma=\gamma_{\rm d}(\Phi_1,\Phi_2)=
\gamma_{\rm c}(\Phi_1,-\Phi_2)$, provided that
$\gamma_{\rm c}(\Phi_1,\Phi_2)=\gamma_{\rm d}(\Phi_1,-\Phi_2)\gg
\tilde \gamma$.
This breakdown of the Onsager relation is a simple manifestation of
the lifting of the time reversal symmetry in the system with time dependent
Hamiltonian.

In the limit of low frequency $\hbar\omega\ll \gamma_{\rm d, c}$, the
contribution from the Cooperon and diffuson parts are described by the
same function, so that the conductance correlation function can be
represented in the form
\begin{equation}
\label{29}
R= \frac{\mls^2 g_{\rm cl}^2}{4\pi^2}\left[\frac{1}{\gamma_{\rm d}^2}
Q\left(\frac{C_0}{\gamma_{\rm d}}, \frac{T}{\gamma_{\rm d}}\right)
+\frac{1}{\gamma_{\rm c}^2}
Q\left(\frac{C_0}{\gamma_{\rm c}}, \frac{T}{\gamma_{\rm c}}\right)
\right],
\end{equation}
so the Onsager relation (\ref{ons1}) holds.
Here,
\begin{eqnarray}
Q(x,y)& =& \int_0^{2\pi}\frac{d\xi d\zeta}{4\pi^2}\int_0^{\infty}
F^2(y\tau)
\\
\nonumber
& \times &
\frac{\exp(-(1+4x\sin^2\xi/2\sin^2\zeta/2)\tau)}
{1+4x\sin^2\xi/2\sin^2\zeta/2} d\tau.
\end{eqnarray}
This expression in the limit of high temperature $T\gg
\gamma_{\rm d, c}$ has an asymptotic behavior 
\begin{equation}
\label{30}
Q(x,y)=\frac{\pi}{3 y}K\left(-4x\right).
\end{equation}
At zero temperature $Q(x,y)$ is given by the expression
\begin{equation}
\label{31}
Q(x,0)=\frac{1}{\pi}\frac{E\left(-4x\right)+(1+4x)K(-4x)}{1+4x},
\end{equation}
where $K(x)$ and $E(x)$ are the elliptic integrals of the first and
second kind respectively
\begin{eqnarray*}
K(x) & = & \int_0^{\pi/2}\frac{d\varphi}{\sqrt{1-x\sin^2\varphi}}\\
E(x) & = & \int_0^{\pi/2}\sqrt{1-x\sin^2\varphi}d\varphi .
\end{eqnarray*}

We conclude that  the conductance fluctuations are suppressed  
by external radiation
even in the limit of the low frequency, see Eq.~(\ref{29}). 
Indeed, during one period of time, the system
goes along a closed loop in the parameter space 
and the contribution to the d.c.-conductance
is effectively determined by the equilibrium conductance, correspondent
to each point of the loop.
The equilibrium conductance fluctuates along this loop.
Thus, the observed d.c.-conductance is already partially
averaged over some realizations of the quantum dot and its
fluctuations decrease. The perturbation strength is related to the
length of the contour in the parameter space and effectively determines
how many different dot's configurations contribute to the
d.c.-conductance. 
Consequently, the stronger perturbation, over the larger number
of the realizations the d.c.-conductance is averaged and the smaller
fluctuations of the d.c.-conductance.
 
This should be contrasted with the suppression of the averaged
magnetoresistance \cite{AAKL,VA}. There, the stationary field does not
do anything because the result is already ensemble averaged.  In order
to suppress the average quantum correction, the field should have
change on the time scale of the order of $1/\gamma_{\rm esc}$, where 
$\gamma_{\rm esc}=\mls N_{\rm ch}/2\pi$ is the escape rate from the
dot. That is why
the effect of the low-frequency radiation on conductance fluctuations
and weak localization corrections are significantly different.
At high frequency $\hbar\omega\gg \gamma_{\rm esc}$ the d.c.-conductance no
longer can be represented in terms of the stationary conductance and
the suppression of both the conductance fluctuations and the weak
localization correction to the conductivity can be interpreted as
dephasing.

{\em Comparison with experiment} ---
Our results still contains an unknown parameter $C_0$ characterizing
the strength of the perturbation. There is a way, however, to present
the results in a form not depending on this parameter, thus
eliminating a need for additional fitting.
Following Ref.~\cite{Huibers}, we represent the parametric dependence
of the weak localization correction $\delta g_{\rm wl}$ versus 
${\rm var}\ \  g$, where $\delta g_{\rm wl}$ is given by\cite{VA}:
\begin{eqnarray}
\label{39}
\delta g_{\rm wl} & = &-\frac{e^2}{\pi\hbar}
\frac{N_{\rm l}N_{\rm r}}{(N_{\rm l}+N_{\rm r})^2}
P\left(\frac{C_0}{\gamma_{\rm esc}},
\frac{\hbar\omega}{\gamma_{\rm esc}}\right)
\\
\label{40}
P(x,z) & = & \int_0^\infty e^{-\xi-x\phi}I_0[x\phi] d\xi, \ \ \
\phi=\xi-\frac{\sin z\xi}{z} .
\end{eqnarray} 
The conductance variance is determined from 
Eq.~(\ref{19}) for broken time-reversal
symmetry $\gamma_{\rm c} \to \infty$. 
Figure~5 shows the parametric dependence for various values
of the parameters $C_0$ and $\omega$ and $T= 10\gamma_{\rm esc}$.

We observe that the shape of the curves depends on the frequency of
external radiation. Particularly, in the limit of low frequency
$\hbar\omega\ll \gamma_{\rm esp}$ the weak localization correction is not
changed by the radiation, while the conductance fluctuations may be
significantly suppressed. At high frequency, $\hbar\omega \gg C_0,\
\gamma_{\rm esc}$ the curves become non-sensitive to the radiation frequency.

The authors of Ref.~\cite{Huibers} found that the
radiation applied to their device produced curves in ${\rm
var} g$ vs. $\delta g_{\rm wl}$ plane identical to the curve
produced by increasing temperature of the device for a wide 
range of frequencies. This observation apparently
demonstrates that the radiation produces the heating of electrons and
the effect of dephasing without heating, see Ref.~\cite{AGA},
is not observed in experiments\cite{Huibers}.
It was also suggested that the main mechanism is
the increase of the temperature in the dot due to the Joule heat by
induced $ac$ source-drain bias.  

\begin{figure}
\centerline{\psfig{figure=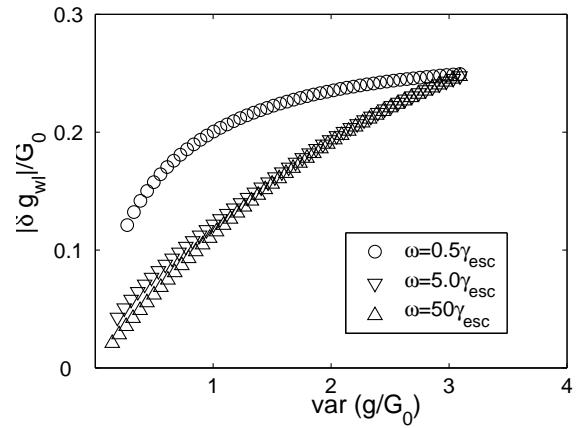,width=7.5cm}}
\caption{
Weak localization correction $\delta g_{\rm wl}$ versus conductance
fluctuations ${\rm var} \ g$ of an open quantum dot with 
$N_{\rm l}=N_{\rm r}\gg 1$ for three values of frequency $\omega$:
$\hbar\omega =0.5\gamma_{\rm esc}$ (o); 
$\hbar\omega=5\gamma_{\rm esc}$ ($\bigtriangledown$);
$\hbar\omega=50 \gamma_{\rm esc}$ ($\bigtriangleup$).
The temperature for all lines was taken $T=10 \gamma_{\rm esc}$.
The amplitude of the field $C_0$ varies from $10^{-2}\gamma_{\rm esc}$
to $10^2\gamma_{\rm esc}$. $G_0=e^2/\pi\hbar$ is the quantum conductance.
}
\end{figure}

Although the present data of Ref.~\cite{Huibers} support the heating
mechanism of suppression of the weak localization correction to the
conductance and the conductance fluctuations, we believe that more
detailed analysis has to be done. According to our theory, see
Eq.~(\ref{17}) and the paragraph below Eq.~(\ref{22.b}), 
({\em i}) mesoscopic fluctuations are
sensitive only to the temperature in the leads, and therefore, the notion
of heating of electrons in the dot responsible for $1/T$ dependence of
the mesoscopic fluctuations is not relevant: if there is a heating,
it manifests itself only through {\em the temperature of the leads};
({\em ii}) high frequency
curves of our theoretical Fig.~5 quantitatively agrees with data on
Fig.~3 of Ref.~\cite{Huibers}, for frequencies $f=1$, $10$ and $25$ GHz.
An exception is the lowest frequency curve ($f=100$ MHz) 
represented in this plot, for
the dot with $\delta_1 = 2.4\mu eV$, $\Nch=2$ corresponds to
$h f/\gamma_{esc}\approx 0.5 $, so according to our Fig.~5 it
should have observable deviations from the high frequency curves,
which is not seen. However, taking into account uncertainty in
determination of the level spacing $\mls$ from the geometrical
area of the dot, this does not unambiguously rule out the microwave
dephasing mechanism.

We believe that the ``smoking gun'' evidence for the mechanism considered
in the present paper is the violation of the Onsager relation (\ref{ons2})
in high frequency regime, $\hbar\omega \gg \gamma_{\rm c,d},\ T$. 
The dependence
of this violation on the amplitude of the field $C_0$ is the main
prediction of the theory. 

{\em Summary}---
We constructed the time dependent random matrix theory to describe the
effect of the non-equilibrium external radiation on conductance
fluctuations of an open quantum dot. The main experimental feature to
reveal such a mechanism is the breakdown of the magnetic symmetry of
the conductance by high-frequency radiation, $\hbar\omega \gg T$, see
Eq.~(\ref{ons2}).

We thank V. Ambegaokar, P. W. Brouwer, V. E. Kravtsov and C. M. Marcus 
for useful discussions. This work was supported by the
Cornell Center for Materials Research, 
under NSF grant no. DMR 0079992 (M.G.V.) and Packard Foundation
Fellowship (I.L.A.).

\widetext


\begin{references}

\bibitem{A1} B.L. Altshuler, Pis'ma Zh. Eksp. Teor. Fiz. {\bf 51}, 530
(1985) [JETP Lett. {\bf 41}, 648 (1985)]. 

\bibitem{LS} P.~A.~Lee and A.~D.~Stone, Phys. Rev. Lett. {\bf 55},
1622 (1985).


\bibitem{A} B.~L.~Altshuler and B.~I.~Shklovskii, Sov. Phys. JETP,
{\bf 64}, 127, (1986).

\bibitem{B} C.W.J. Beenakker, Rev. Mod. Phys, {\bf 69}, 731, (1997).

\bibitem{AK} B. L. Altshuler and D. E. Khmelnitskii, JETP Lett. {\bf
42}, 359 (1985).
Also see P. A. Lee,
Fukuyama, A.D Stone Phys. Rev. B {\bf 35}, 1039 (1985).


\bibitem{H1} A.~G.~Huibers {\it et. al.} Phys. Rev. Lett. {\bf 81},
200 (1998).

\bibitem{Huibers} A.~G.~Huibers, J.~A.~Folk, S.~R.~Patel,
C.~M.~Marcus, C.~I.~Duru\"oz and J.~S.~Harris, Phys.~Rev.~Lett.,
{\bf 83}, 5090, (1999).

\bibitem{AAKL} B.L. Altshuler, {\em et. al.}, in
{\it Quantum Theory of Solids}, (Mir, Moscow, 1982).

\bibitem{BM1} 
H.U.~Baranger, P.A.~Mello, Phys. Rev. B, {\bf 51}, 4703, (1995);
P.W.~Brouwer and C.W.J.~Beenakker, 
{\em ibid}., {\bf 51}, 7739, (1995);
{\em ibid}., {\bf 55}, 4695, (1997);
I.L.~Aleiner and A.I.~Larkin. {\em ibid}., {\bf 54},
14423 (1996);
E. McCann and I.V. Lerner, {\em ibid}.,
{\bf 57}, 7219 (1998).

\bibitem{VA} M.G.~Vavilov and I.L.~Aleiner, Phys. Rev. B, {\bf 60},
R16311, (1999). 

\bibitem{VAA} M.~G.~Vavilov, V.~Ambegaokar, and I.~L.~Aleiner,
cond-mat/0008469. 

\bibitem{Brouwer-ac} P.~W.~Brouwer, Phys. Rev. B {\bf 63}, 121303 (2001).

\bibitem{Y} The same conclusion for the conductance fluctuations of
open quantum dots was reached  by V.~I.~Yudson,
E.~Kanzieper, and V.~E.~Kravtsov, cond-mat/0012200. We do not quite
understand reasoning of this paper, since the electroneutrality
condition is not invoked there. Nevertheless, the answer is 
correct.

\bibitem{WK} X.-B. Wang and V.~E.~Kravtsov, cond-mat/0008193.

\bibitem{ABG} I. L. Aleiner, P. W. Brouwer, and L. I. Glazman, 
cond-mat/0103008.

\bibitem{Brouwer} P.W. Brouwer and I.L. Aleiner, Phys. Rev. Lett., 
{\bf 82}, 390 (1999).

\bibitem{AGD} A. A.~Abrikosov, L. P.~Gorkov, I. E.~Dzyaloshinskii, 
{\it Methods 
of Quantum Field Theory in Statistical Physics}, (Prentice--Hall, Englewood 
Cliffs, NJ, 1963).

\bibitem{fn1} Note that $C_0$ is defined in ref.~\cite{VA} with a
different numerical factor.

\bibitem{SA} B.D.~Simons and B.L.~Altshuler, Phys. Rev. Lett., {\bf 70}, 4063,
(1993); B.L.~Altshuler and B.D.~Simons, in {\it Mesoscopic Quantum Physics},
eds. E.~Akkermans {\it et. al}, (Elsevier, 1995). 

\bibitem{Buttikker} M. B\" uttikker, H. Thomas, and A. Pretre, Z. Phys. B
{\bf 94}, 133 (1994).

\bibitem{Onsager} L. Onsager, Phys. Rev. {\bf 38}, 2265 (1931); 
M. B\" uttikker, Phys. Rev. Lett. {\bf 57}, 1761 (1986).

\bibitem{AGA}  B.~L.~Altshuler,  M.E.~Gershenzon, and I.L.~Aleiner,
Physica E {\bf 3}, 58 (1999). 

\end{references}
\end{document}